\documentclass[12pt,a4paper]{article}

\usepackage{soul}        

\usepackage{amsmath,amssymb,epsf,epsfig}
\usepackage{array}
\usepackage{fancybox}

\usepackage[usenames]{color}
\usepackage{amsfonts,bm}
\usepackage{graphicx}
\usepackage{enumerate}

\usepackage{parskip}     
\usepackage[
      colorlinks=true,
      linkcolor=blue,
      urlcolor=blue,
      filecolor=blue,
      citecolor=blue,
      pdfstartview=FitH,
      pdftitle={},
      pdfauthor={},
      pdfsubject={},
      pdfkeywords={},
      pdfpagemode=UseNone,
      bookmarksopen=true
      ]{hyperref}
\usepackage[all]{hypcap}     

\numberwithin{equation}{section}

\newcommand{\be}{\begin{equation}}
\newcommand{\ee}{\end{equation}}
\newcommand{\bea}{\begin{eqnarray}\displaystyle}
\newcommand{\eea}{\end{eqnarray}}

\def\beq{\begin{equation}}
\def\eeq{\end{equation}}
\def\beqa{\begin{eqnarray}}
\def\eeqa{\end{eqnarray}}
\def\bet{\begin{tabular}}
\def\eet{\end{tabular}}
\def\bs{\begin{split}}
\def\es{\end{split}}

\def\one{{\hbox{\kern+.5mm 1\kern-.8mm l}}}
\def\zero{{\hbox{0\kern-1.5mm 0}}}

\definecolor{orange}{rgb}{1,0.5,0}

\def\ii{{\rm i}}

\newcommand{\bean}{\begin{eqnarray*}}
\newcommand{\eean}{\end{eqnarray*}}

\begin{document}

\hfill \hbox{DFPD-12-TH-01} 

\vspace{-0.5cm}
\hfill \hbox{QMUL-PH-12-02}

\vspace{2cm}

\centerline{
\LARGE{ \textsc{Adding new hair to the 3-charge black ring}} }



\vspace{1.5cm}

\centerline{    
  \textsc{ Stefano Giusto$^{1,2}$, ~Rodolfo Russo$^3$}  }

\vspace{0.5cm}

\begin{center}
$^1\,${Dipartimento di Fisica ``Galileo Galilei'',\\
Universit\`a di Padova,\\ Via Marzolo 8, 35131 Padova, Italy\\
}
\end{center}

\vspace{0.1cm}

\begin{center}
$^2\,${INFN, Sezione di Padova,\\
Via Marzolo 8, 35131, Padova, Italy}
\end{center}
\vspace{0.1cm}

\begin{center}
$^3\,${Queen Mary University of London,\\
Centre for Research in String Theory, School of Physics\\
Mile End Road, London E1 4NS, UK\\
}
\end{center}

\vspace{0.2cm}

\begin{center}
{\small stefano.giusto@pd.infn.it, ~~r.russo@qmul.ac.uk}
\end{center}

\vspace{1cm}

\centerline{ 
 \textsc{ Abstract}}

\vspace{0.2cm} {\small Motivated by the string theory analysis
  of~\cite{Giusto:2011fy}, we construct a class of $1/8$-BPS solutions
  of type IIB supergravity compactified on $S^1 \times T^4$. In this
  duality frame our ansatz allows for a non-trivial NS-NS B-field
  which has been usually set to zero in previous studies of $1/8$-BPS
  geometries. We provide a M-theory description of these new
  geometries and show that they can be interpreted as the lift of
  solutions of the ${\cal N}=2$ 5D supergravity with three vector
  multiplets and whose scalar manifold is the symmetric space $SO(1,1)
  \otimes (SO(1,2)/SO(2))$. Finally we show that the non-minimal 5D
  black rings provide an explicit example of solutions falling in this
  ansatz. In particular we point out the existence of a black ring
  that has an extra dipole charge with respect to the solutions of
  the STU-model. In the near-horizon limit, this ring has an AdS$_3
  \times S^3$ geometry with the same radius as the one of the 3-charge
  black hole and thus its microstates should belong to the usual D1-D5
  CFT.}
        
\thispagestyle{empty}

\vfill
\eject

\setcounter{page}{1}

\section{Introduction} \label{sec:introduction}

D-brane configurations in type II string theories can be described
both from a gravitational and a microscopic ({\em i.e.} conformal
field theory) point of view, see for
instance~\cite{Duff:1994an,Polchinski:1996fm} for reviews focusing on
the first and the second aspect respectively. A link between these two
descriptions is provided by the conserved quantities, such as the
charges and the energy of the configuration under analysis. These
quantities can be easily extracted from the supergravity solution by
looking at the large distance decay of the relevant fields; for
instance the energy is obtained from the time component of the metric
$g_{00}$ and, in the D-brane case, the charge from the RR fields. In
the microscopic description, where D-branes are defined as the place
where open strings can end, these conserved quantities are obtained
from 1-point CFT correlators where the string world-sheet has the
topology of the disk~\cite{DiVecchia:1997pr}.

More recently~\cite{Giusto:2009qq,Black:2010uq}, it was showed for
$1/4$-BPS configurations in type II supergravity that this connection
between classical solutions and disk CFT correlators holds also for
higher order terms in the large distance expansion, even if they
capture dipole instead of conserved charges. In~\cite{Giusto:2011fy},
the CFT approach was used to study a class of $1/8$-BPS D-brane
configurations in type IIB string theory on $R^{1,4} \times S^1\times
T^4$. From the space-time point of view the brane configuration under
analysis was a bound state of a D1 and a D5-brane each wrapped $n_w$
times on the $S^1$ and oscillating in the common $R^4$ Dirichlet
directions. When this oscillation is described by a purely left (or
right) moving null-like wave the system preserves 4 of the original 32
supersymmetries. By studying the disk correlators with the insertion
of a closed string vertex corresponding to the massless bosonic
degrees of freedom, an asymptotic supergravity solution was derived
including dipole and quadrupole terms in the $1/r$
expansion. In~\cite{Giusto:2011fy} it was also checked, in
perturbation theory, that the solution arising from the CFT analysis
was indeed $1/8$-BPS.

In this paper, we show that the asymptotic solution
of~\cite{Giusto:2011fy} is an example of a class of classical
configurations which solve the full non-linear equations of
supergravity, {\em i.e.} the supersymmetry variations with four
independent parameters and the equations of motion. The structure of
these solutions is similar to that of the configurations
studied in~\cite{Kanitscheider:2007wq}, and this provided a crucial
guidance in the non-linear generalization of the results derived from
string correlators. In many respects, the case discussed here can be
seen as a direct generalization of the results
of~\cite{standardBPS}, but, as suggested
by the string analysis, the new supergravity ansatz depends on an
additional scalar and vector in the uncompact $R^{1,4}$. Since the
original 10D ansatz of~\cite{standardBPS} 
could be interpreted as the lift of a $1/2$-BPS
solution in a 5D ungauged supergravity with two vector multiplets (the
STU-model), it is natural to suspect that the generalization discussed
here is a solution of an extended 5D supergravity with an extra vector
multiplet. We present a 11D lift of the IIB ansatz which makes the
embedding of this ansatz in such a 5D supergravity manifest (this also
provides an alternative and maybe simpler way to check the equations
of motions).

In order to provide a concrete solution which falls in the new class
discussed in this paper, we use the non-minimal black rings studied
in~\cite{BR}, which are solutions of the ${\cal N}=2$ supergravity
with $n$ vector multiplets. We focus on the case $n=3$, which has one
more charge and dipole with the respect to the original black ring of
the STU-model~\cite{Elvang:2004rt}. We provide an embedding of the
$n=3$ black ring in type IIB string theory, which is a first step
towards understanding the ring microstates. In M-theory frame, there does not
seem to be any compelling reason to concentrate on solutions with $n=3$ vector multiplets.
However, as it emerges from our analysis, this subclass of solutions is very natural from a type IIB perspective because three is the maximum number of vector fields
one can turn on while still preserving the $SO(4)$ rotations of the $T^4$ 
directions in the IIB duality frame. It is not obvious to us
that the $n=3$ black ring solution can be derived from the one
of the STU model by using some generating solution
techniques. For instance, U-duality transformations do not seem to be
enough to connect the two types of black ring solutions in 10D.

The paper has the following structure. In Section~\ref{sec:solution}
we start from the perturbative solution of~\cite{Giusto:2011fy} and
motivate the form of the full-nonlinear solution in a type IIB frame
where the charges correspond to momentum, D1 and D5 branes. The
solution is given in terms of 4 scalar functions and 5 vectors defined
on a hyper-Kahler metric for the $R^4$ corresponding to the Dirichlet
directions. In Section~\ref{sec:mtheorylift} we provide a M-theory
lift of this class of solutions which makes it manifest how to perform
the embedding in a ${\cal N}=2$ 5D ungauged
supergravity~\cite{Gunaydin:1983bi}, where the scalar manifold is the
symmetric space $SO(1,1) \otimes (SO(1,2)/SO(2))$. It is interesting
that the lift from type IIB to M-theory used to embed the ansatz
of~\cite{standardBPS} in the ${\cal N}=2$
5D STU-model does not work for the case described in this paper and so
we are required to consider a more involved U-duality transformation
to get a simple 11D configuration. In Section~\ref{sec:BR}, we
consider a black ring which gives an explicit example of a solution
falling in the ansatz discussed. This configuration has a non-trivial
horizon area which, as it happened for the original black ring, is
given by the quartic invariant of $E_7$~\cite{Cremmer:1979up}. In
Section~\ref{daop} we briefly comment on how to embed the black ring
solution under analysis in AdS$_3 \times S^3$. This is possible by
taking a near horizon limit on the solution written in the type IIB
duality frame; then one can apply the usual dictionary and use the AdS
radius to derive the central charge of the CFT that should describe
the microstates of the black ring. In particular, we can choose a ring
that has the same charges and dipoles of the perturbative string
solution of~\cite{Giusto:2011fy} and starts to differ from it only at
the level of quadrupole corrections. In this case the AdS asymptotic
is the same as the one of the D1-D5 case and so the dual description
should be given by the same D1-D5 CFT describing the Strominger-Vafa
and BMPV black holes. In the spirit of the analysis by Mathur et
al.~\cite{fuzzball},
we interpret this as evidence for the existence of a class of
microstate solutions that have the same dipole terms of the black
ring, but then have also quadrupole terms, as predicted by the
microscopic D-brane analysis of~\cite{Giusto:2011fy}. In
Appendix~\ref{sec:dualities} we collect the equations of motions, the
supersymmetry variations and the duality rules needed to reproduce the
results presented in the main text.

\section{A new $1/8$-BPS ansatz in type IIB supergravity} \label{sec:solution}

Let us start by recalling the type IIB supergravity solutions
belonging to the class discussed
in~\cite{standardBPS}. They are
characterised by three charges corresponding to momentum, D1 and D5
branes, and three dipoles related to KK monopoles, D1 and D5
branes. We indicate the $S^1$ where all branes are wrapped by the
coordinate $y$ and the D5 branes wrap also the $T^4$ describing the
remaining four compact directions. The uncompact space is
topologically $R^{1,4}$, where the $R^4$ part is described by a 4D
hyper-Kahler metric $ds_4^2$. Then the 10D (string frame) metric is
given by
\bea\label{bwg}
ds^2&=& \frac{1}{\sqrt{Z_1 Z_2}}\,\Bigl[-\frac{1}{Z_3}\,d\hat t^2 +
Z_3\,d\hat y^2\Bigr] + \sqrt{Z_1 Z_2} \,ds^2_4 +
\sqrt{\frac{Z_1}{Z_2}}\, ds^2_{T^4}\,,\\ \label{tyviel}
d\hat t &=& dt + k \,,\quad d\hat y = dy+dt -\frac{dt +k}{Z_3}+a_3\,,
\eea 
where $Z_i$, with $i=1,2,3$, are scalar functions, while $a_3$ and $k$
are 1-forms in $R^4$. $ds^2_{T^4}$ denotes the metric on $T^4$, for which we simply take  
$ds^2_{T^4}= \sum_{i=1}^4 dz_i^2$. The other non-trivial fields are the dilaton and
the RR 2-form
\bea\label{bwof}
C^{(2)}&=& -\frac{1}{Z_1}\,d\hat t\wedge d\hat y + a_1\wedge\Bigl(d\hat y + \frac{d\hat t}{Z_3}\Bigr) +\gamma_2\,,~~~~{\rm e}^{2\phi} =\frac{Z_1}{Z_2}\,,
\eea 
where $\gamma_2$ is a 2-form and $a_1$ is another 1-form in $R^4$. In
order to preserve four supercharges and solve the type IIB
supergravity equation of motions, the scalars and the forms appearing
in the ansatz above have to satisfy a set of differential equations in
$R^4$. The 1-forms $a_{1,3}$ have a self-dual field strength with the
respect to the $ds_4^2$ metric. Also we can define a new form $a_2$,
with the same property, starting from the self-dual part of the $k$
\be\label{eq2bw} 
Z_2\,da_2 \equiv dk+*_4 dk - Z_1\,da_1-Z_3\,da_3 \,.
\ee 
Thus in summary we have $*_4 da_i = da_i$ for $i=1,2,3$ and the
scalars $Z_i$ have to satisfy
\be\label{eq4bw} 
d *_4 d Z_1 = -
da_2\wedge da_3\,,\,\,\, d *_4 d Z_2 = - da_1\wedge da_3\,,\,\,\,\, d
*_4 d Z_3 = -da_1\wedge da_2 \,, 
\ee 
Finally the field strength of the 2-form $\gamma_2$ is given in terms of
the $Z_i$ and $a_i$ by
\be\label{gammabw} 
d\gamma_2 =*_4 d Z_2 + a_1 \wedge da_3\,.  
\ee
Notice that~\eqref{eq4bw} ensures that the r.h.s. of this equation is
closed and can be written, at least locally, as the differential of a
2-form $\gamma$.

The main new feature of the asymptotic solution
in~\cite{Giusto:2011fy} is that also the other massless type IIB
fields are non-trivial. In particular, at order $1/r^3$ in the large
distance expansion, the NS-NS B-field components $B_{ti}$ and
$B_{yi}$, where $i$ is in the uncompact $R^4$, are non-zero. Then at
order $1/r^4$ also the 10D axion, the $B_{ty}$, $B_{ij}$ components
and the 5-form RR field strength are non-trivial. It is interesting
that, in the string results, the two new $1/r^3$ terms are fixed by
the same $R^4$ vector, while all the new $1/r^4$ terms are determined
by a single scalar. By following the structure of the ansatz
summarised above, we will refer to the new scalar and the new vector
with self-dual field strength as $Z_4$ and $a_4$ respectively. Even if
the supergravity asymptotic solution arising from string correlators
looks very complicated, as many components are non-trivial, the new
terms are actually rather similar to those appearing in a class of
solutions discussed in~\cite{Kanitscheider:2007wq}. These are
$1/4$-BPS solutions with only D1 and D5 brane charges which were
derived by a U-duality from the solution describing a fundamental
string with a left moving wave whose plane of oscillation has one
direction in the compact space $T^4$. In particular, we are interested
in the case where, in the D1/D5 frame, the solution still preserves
the $SO(4)$ rotations of the $T^4$ directions in the IIB duality
frame. This configuration was analysed from the string point of view
in~\cite{Giusto:2009qq}. In the $1/4$-BPS case all the fields that in
our $1/8$-BPS case are determined by $Z_4$ appear at order $1/r^3$. In
the notation of~\cite{Giusto:2009qq}, these fields were determined by
the function ${\cal A}$. The natural guess is that our new scalar
$Z_4$ should play in the $1/8$-BPS case under consideration exactly
the same role of ${\cal A}$. This provides a guide to complete the
non-linear dependence on $Z_4$ of the new solution. So we introduce
the combination
\begin{equation}
  \label{eq:alpha}
  \alpha = \Bigl(1-\frac{Z^2_4}{Z_1 Z_2}\Bigr)^{-1}\,,
\end{equation}
which should play the same role as the ratio $\tilde{H}_1/H_1$ in the
$1/4$-BPS solution of~\cite{Kanitscheider:2007wq}. The other natural
guess is that the self-duality property of $da_4$ holds exactly
including all the non-linear correction of the 4D hyper-Kahler
metric. 

Thus by combining the previous supergravity analysis and the string
theory results of~\cite{Giusto:2011fy}, we are led to the following
ansatz 
\begin{align}
ds^2&= \frac{\alpha}{\sqrt{Z_1 Z_2}}\,\Bigl[-\frac{1}{Z_3}\,d\hat t^2
+ Z_3\,d\hat y^2\Bigr] + \sqrt{Z_1 Z_2} \,ds^2_4 +
\sqrt{\frac{Z_1}{Z_2}}\, ds^2_{T^4}\,,\nonumber
\\ 
\nonumber
d\hat t &= dt + k \,,\quad d\hat y = dy+dt -\frac{dt +k}{Z_3}+a_3\,,\\
\label{na}
{\rm e}^{2\phi} &= \alpha\,\frac{Z_1}{Z_2}\,,\quad C^{(0)} =
\frac{Z_4}{Z_1}\,,
\\ \nonumber
C^{(2)}&= -\frac{\alpha}{Z_1}\,d\hat t\wedge d\hat y +
a_1\wedge\Bigl(d\hat y + \frac{d\hat t}{Z_3}\Bigr)
+\gamma_2\,,
\displaybreak \\ \nonumber
B^{(2)}&= -\frac{\alpha Z_4}{Z_1 Z_2}\,d\hat t\wedge d\hat y
+ a_4\wedge\Bigl(d\hat y + \frac{d\hat t}{Z_3}\Bigr)
+\delta_2\,, 
\\ \nonumber
F^{(5)} &= d\Bigl(\frac{Z_4}{Z_2}\Bigr) \wedge dz^4 +
\alpha\,\frac{Z_2}{Z_1} *_4 d \Bigl(\frac{Z_4}{Z_2}\Bigr) \wedge d\hat
t \wedge d\hat y\,.
\end{align}
We denote by $dz^4$ the volume form of $T^4$: $dz^4 = dz_1\wedge\ldots\wedge dz_4$.
The 2-form $\delta_2$ plays the same role as the $\gamma_2$ in~\eqref{bwof}
\be\label{delta}
d\delta_2 =*_4 d Z_4 + a_4 \wedge da_3\,.
\ee
One can check that the ansatz~\eqref{na} preserves four supercharges
and solves the type IIB equations of motion, provided that the
definition of $a_2$ in~\eqref{eq2bw} is slightly modified 
\be\label{eq2na} 
Z_2\,da_2 \equiv dk+*_4 dk - Z_1\,da_1-Z_3\,da_3 + 2 Z_4\,da_4
\ee 
and the equation of the scalars take the form
\begin{align}\label{eq4na} 
&d *_4 d Z_1 = - da_2\wedge da_3\,,\,&\,\, 
d *_4 d Z_2 = - da_1\wedge da_3\,,\\ \nonumber
&d *_4 d Z_3 = -da_1\wedge da_2 + da_4 \wedge da_4 \,, \,&\,\,
d *_4 d Z_4 = -da_3\wedge da_4 \,.
\end{align}
We will see that Eqs.~\eqref{eq2na} and~\eqref{eq4na} implies the
equations of motions of a ${\cal N}=2$  5D ungauged supergravity with
three vector multiplets for the case static solutions.

\section{M-theory lift and ${\cal N}=2$ truncation} \label{sec:mtheorylift}

The standard approach to lift~\eqref{bwg}--\eqref{bwof} to M-theory is
to take three T-dualities, one along the $S^1$ direction $y$ where the
D-branes are wrapped, and the remaining two in the $T^4$. By trying
the same approach starting from~\eqref{na}, we get a rather
complicated 11D solution which we could not truncate to any 5D
supergravity. The main problem is that the new 1-form $a_4$ in the
$B$-field~\eqref{na} appears in the 11D metric and is not on the same
footing as the other forms which appear in the potential
$A^{(3)}$. This suggests that the solution to this problem is to look
for a different duality chain. We indicate with $S$ a type IIB
S-duality, with $T_{ab}$ a pair of T-dualities along the Cartesian
coordinates $z_a,z_b$, $a,b=1,\ldots,4$ of the (square) $T^4$, and
$T_y$ a T-duality along the $S^1$ parametrised by $y$. Then the chain
of dualities we are going to consider is
\be\label{dc}
\begin{array}{ccccccccc}
T_{12}&\to& S &\to& T_{13} &\to& T_y &\to& \mathrm{11D~lift} \\
IIB & & IIB & & IIB & & IIA & & M-th
\end{array}
\ee
It is not difficult to follow the fate of the 1-forms $a_I$
in~\eqref{na} and realise that these dualities bring the type IIB
ansatz to a nice M-theory frame, where all $a_I$ appear as components
the 3-form potential $A^{(3)}$. By using the rules summarised in the
Appendix~\ref{sec:dualities}, we can follow how the ansatz~\eqref{na}
transforms under the duality chain in~\eqref{dc}. In order to carry
out this computation, it is convenient to use~\eqref{fieldstrengths} and derive
the potential $C^{(4)}$ for the self-dual field strength in~\eqref{na}
\be
C^{(4)} = \frac{{Z_4}}{Z_2}\,dz^4-\frac{\alpha\,{Z_4}}{Z_1 Z_2}\,\gamma_2\wedge d\hat t\wedge d\hat y + x_3\wedge\Bigl(d\hat y + \frac{d\hat t}{Z_3}\Bigr)\,,
\ee
where 
\be
dx_3 = da_4\wedge \gamma_2-a_1\wedge (d\delta_2 - a_4\wedge da_3)\,.
\ee
It is also convenient to work in a `democratic' formalism where both
the R-R field strengths and their Hodge dual are explicitly written in
the solution. Clearly the high-degree $n$-forms, with $n>5$, contain
redundant information, but after some T-dualities they can contribute
to a low degree form with $n\leq 5$. For our specific case only $C^{(6)}$ is needed; it is defined via~\eqref{dualfields} and~\eqref{fieldstrengths} and it is given by
\be
C^{(6)}=\Bigl[-\frac{\alpha}{Z_2}\,d\hat t\wedge d\hat y + a_2\wedge\Bigl(d\hat y + \frac{d\hat t}{Z_3}\Bigr) +\gamma_1\Bigr]\wedge dz^4 - \frac{\alpha\,Z_4}{Z_1 Z_2}\gamma_2\wedge \gamma_2\wedge d\hat t\wedge d\hat y\,,
\ee
with
\be
d\gamma_1 =*_4 d Z_1 + a_2 \wedge da_3\,.
\ee
The result for the 11D metric is
\bea \label{mm}
ds^2_{11} &=& -\Bigl(\frac{\alpha}{Z_1 Z_2 Z_3}\Bigr)^{2/3}\,d\hat t^2
+\Bigl(\frac{Z_1 Z_2 Z_3}{\alpha}\Bigr)^{1/3}\,ds^2_4
\\ \nonumber 
&+&\alpha^{2/3}\,(Z_1 Z_2 Z_3)^{1/3}\,\Bigl[\frac{dw_1\,d\bar
  w_1}{Z_1} +\frac{dw_2\,d\bar w_2}{Z_2}+\frac{dw_3\,d\bar
  w_3}{\alpha\,Z_3}
\\ \nonumber 
&+& \frac{Z_4}{Z_1 Z_2}\,(dw_1\,d\bar w_2 +dw_2\,d\bar w_1) \Bigr]\,,
\eea
where the complex coordinates $w_{1,2}$ parametrise the $T^4$ that was
already present in the type IIB setup, while $w_3$ parametrises an
extra $T^2$ obtained by combining the $y$ direction together with the
M-theory circle $z$
\be\label{ccw}
w_1 \equiv z_2-\ii\,z_3\,,\quad w_2 \equiv z_1+\ii\,z_4\,,\quad w_3 \equiv
y+\ii\,z\,. 
\ee
Notice that the volume of the $T^6$ defined by the second and third
line of~\eqref{mm} is independent of all functions $Z_I$ and is simply
equal to one. For the 11D 3-form potential we obtain
\bea\label{3fm}
A^{(3)}&=& \Bigl(-\frac{\alpha \,d\hat t}{Z_1} + a_1\Bigr)\wedge
\frac{dw_1\wedge d\bar w_1}{-2\,\ii}+ \Bigl(-\frac{\alpha\, d\hat t}{Z_2}
+ a_2\Bigr)\wedge \frac{dw_2\wedge d\bar w_2}{-2\,\ii}
\\ \nonumber
&+& \Bigl(-\frac{d\hat t}{Z_3} + a_3\Bigr)\wedge \frac{dw_3\wedge
  d\bar w_3}{-2\,\ii}+  \Bigl(-\frac{\alpha\,Z_4}{Z_1Z_2}\,d\hat
t + a_4 \Bigr)\wedge \frac{dw_1\wedge d\bar w_2 + d w_2 \wedge d\bar
  w_1}{-2\,\ii}\,. 
\eea
The non-trivial components in~\eqref{mm} and~\eqref{3fm} are the only
ones that are invariant under the two continuous transformations
\be\label{tratro1}
(w_1 \to {\rm e}^{\ii \theta} w_1\,,~ w_2 \to {\rm e}^{\ii \theta} w_2)
~~~~\mbox{and} ~~~~~ w_3 \to {\rm e}^{\ii \phi} w_3\,,
\ee
and the discrete transformation
\be\label{tratro2}
(A^{(3)} \to - A^{(3)}\;,~ w_1 \leftrightarrow \bar{w}_1\;,~ w_2
\leftrightarrow \bar{w}_2 \;,~ w_3 \leftrightarrow \bar{w}_3) \,.
\ee
These are symmetries of the original 11D supergravity and thus the
restriction to the fields that are invariant under~\eqref{tratro1}
and~\eqref{tratro2} provides a consistent truncation. The
result~\eqref{3fm} for $A^{(3)}$ suggests to introduce 
\be\label{cycles}
J_i = \frac{dw_i\wedge d\bar w_i}{-2\,\ii}\;,~~i=1,2,3~~~~~\mbox{and}~~~~
J_4 = \frac{dw_1\wedge d\bar w_2 + d w_2 \wedge d\bar w_1}{-2\,\ii}\,.
\ee
These are $(1,1)$-forms with the respect to the complex
structure~\eqref{ccw}. So we can
follow~\cite{Mtheory} and list the ${\cal
  N}=2$ vector multiplets obtained by reducing the 11D ansatz on $T^6$
down to 5D. The intersection numbers of the $J_I$ will play an
important role
\be
J_I\wedge J_J \wedge J_K = \frac{1}{6}\, C_{IJK}\, J\wedge J \wedge J
\,,
\ee
where the Kahler form $J$ is given in~\eqref{scalre}. The $C$'s are
clearly fully symmetric under the exchange of their indices and, in
our case, $C_{123}$ and all its permutations are equal to $1$, while
the only non-trivial coupling involving $I=4$ is $C_{344}$
\be\label{inu}
C_{123}= 1\,,\quad C_{344} = -2\,.
\ee
The scalars of the vector multiplets can be read by decomposing $J$
along the $J_I$
\be\label{scalre}
J= \ii\, h_{i \bar{k}}\, dw^i \wedge d\bar{w}^k = \sum_I t^I J_I~,
~~~\mbox{with}~~~~~
ds^2_{T_6} = h_{i \bar{k}}\, dw^i d\bar{w}^k \,.
\ee
In writing the Kahler form $J$ for our case we keep only the
components that are non-trivial in~\eqref{mm} and obtain
\be\label{Jde}
J = \alpha^{2/3}\,(Z_1 Z_2
Z_3)^{1/3}\Bigl[\frac{1}{Z_1}\,J_1+\frac{1}{Z_2}\,J_2+\frac{1}{\alpha\,Z_3}\,J_3+\frac{Z_4}{ Z_1Z_2}\,J_4\Bigr]\,.
\ee
This yields the moduli coordinate $t^I$. Since, as noticed above, the
total volume of the compact space is just one in our case the $t^I$
automatically satisfy the relation $t^It^J t^K C_{IJK}/6=1$. By
using~\eqref{inu} for the intersection numbers for our $J_I$
in~\eqref{cycles}, we can write the constraint on the moduli
coordinates as 
\be\label{tttC}
\mathcal{V} = \frac{1}{6} t^I t^J t^K \, C_{IJK} =t^1 t^2 t^3 - t^3
(t^4)^2 =1 \,. 
\ee
This constraint defines the symmetric space $SO(1,1) \otimes
(SO(1,2)/SO(2))$, which appears in one of the possible ${\cal N}=2$ 5D
truncation of the maximal ungauged supergravity.

In a similar way we can decompose $A^{(3)}$ along the $J_I$ and read
the vectors $A_I$ obtained in the reduction $A^{(3)} = \sum_I
A_I\,J_I$. In our case we have
\be\label{vecre}
A^{(3)} = 
\sum_{c=1}^2 \left(-\frac{\alpha\, d\hat t}{Z_c} + a_c\right) J_c + 
\left(-\frac{d\hat t}{Z_3} + a_3\right) J_3 + 
\left(-\frac{\alpha Z_4}{Z_1 Z_2}\,d\hat t + a_4\right)
J_4\,.
\ee
Contrary to what happens in the expansion of the Kahler
form~\eqref{Jde}, in the case of $A^{(3)}$ there is no constraint
relating the components $A_c$. This means that the number of the 5D
vectors is equal to the number $h_{11}$ of $(1,1)$-forms, while the
number of independent scalars obtained from $J$ is just
$h_{11}-1$. This field content fits in $h_{11}-1$ ${\cal N}=2$ vector
multiplets, while the additional vector in~\eqref{vecre} is part of
the graviton multiplet.

In our case~\eqref{cycles} $h_{11}=4$, so we have three vector
multiplets. The explicit form of the intersection numbers~\eqref{inu}
fixes completely the 5D action
\be
\int d^5x \left[\sqrt{|g_5|} R-\frac{Q_{IJ}}{4}\, *_5
  F^I \! \wedge \! F^J - \frac{Q_{IJ}}{2}\,*_5 d t^I\! \wedge \! d t^J -
\frac{C_{IJK}}{24}\,F^I\! \wedge \! F^J\! \wedge \! A^K\right]\,, 
\ee
where the quadratic form $Q$ in the kinetic terms is 
\be
Q_{IJ} = -\frac{\partial}{\partial t^I}\,\frac{\partial}{\partial
  t^J}\log \mathcal{V}\, |_{\mathcal{V}=1}\,,
\ee
and $F^I= d A^I$.
With the identifications~\eqref{scalre} and~\eqref{vecre} between the
fields in the ${\cal N}=2$ supergravity, one can check that the 5D
equations of motions are implied by the self-duality property $*_4
da_I = da_I$ together with Eqs.~\eqref{eq2na},~\eqref{eq4na}. For instance the
equations of motion for the vectors are
\be
d (Q_{IJ} \,*_5 F^J) =\frac{C_{IJK}}{2}\,F^J\wedge F^K\,,
\ee
and its magnetic components are equivalent to~\eqref{eq4na}. Thus we
can see the solutions falling in the ansatz~\eqref{mm} and~\eqref{3fm}
as the 11D lift of 5D solutions of an ${\cal N}=2$ supergravity.

\section{A black ring solution} \label{sec:BR}

In~\eqref{na} the NS-NS and the R-R 2-forms have a very similar
structure, where the only qualitative difference is that $\delta_2$ is
related to the scalar $Z_4$ appearing the $dt\wedge dy$ term of
$B^{(2)}$, while the corresponding components for $C^{(2)}$ are
completely independent. So this ansatz allows for solutions with equal,
but non-trivial, charges for a fundamental string and a NS5-brane. The
simplest possible example of a solution of this type is probably a
black hole characterised by four parameters $Q_I$ determining its
charges. It is natural to expect that there exists also a class of
black ring solutions with four parameters fixing the charges and four
parameters related to dipoles. Such a black ring solution has indeed already
appeared as a particular case of solutions in~\cite{BR},
that constructed black ring solutions in 5D $\mathcal{N}=2$ supergravity
coupled to an arbitrary number of vector multiplets. Here we emphasize the
11D interpretation of this solution and, more importantly, the fact that it is U-dual
to a IIB solution belonging to the ansatz that emerges from the string theory computation
of~\cite{Giusto:2011fy}. We believe that this black ring solution cannot be derived
 by using some generating solution techniques from the black
ring with three charges and three dipoles.

Following the by now well established construction of BPS solutions with 
two axial symmetries, we choose the 4D part
$ds_4^2$ in~\eqref{na} to be simply the flat Euclidean metric and rewrite it in the Gibbons-Hawking form:
\be\label{ghm}
ds^2_4 = V^{-1} (d\tau+A)^2 + V\,ds^2_3\,,\quad *_3 d A = dV\,,
\ee
where $\tau$ is a linear combination of the two axial isometries of
two orthogonal planes in $R^4$ and $ds^2_3$ is the flat 3-dimensional
metric
\be\label{3df}
ds^2_3 = dr^2 + r^2 \,d\theta^2 + r^2 \,\sin^2\theta\, d\phi^2\,.
\ee
Since we are focusing on a flat 4D metric, the Gibbons-Hawking
potentials are 
\be
V = \frac{1}{r}\,,~~~~
A = \cos\theta d\phi~.
\ee
For configurations preserving a $U(1)\times U(1)$ symmetry, such as
the black rings, we can express $k$, the $a_I$'s and the $Z_I$'s of
the ansatz~\eqref{na} in terms of harmonic functions $K_I$, $L_I$ and
$M$ on $R^3$
\bea \label{f2t}
&&a_I = \frac{K_I}{V}\,(d\tau+A) + {\bar a}_I\,,\quad *_3 d{\bar a}_I
= - d K_I\,, \quad (I=1,2,3,4)\,,\\
&&Z_1 = L_1 + \frac{K_2\,K_3}{V}\,,\quad Z_2 = L_2 +
\frac{K_1\,K_3}{V}\,, \nonumber \\
&& \quad Z_3 = L_3 +\frac{K_1\,K_2}{V} -
\frac{K_4^2}{V}\,,\quad Z_4 = L_4 + \frac{K_3\,K_4}{V}\,,\nonumber\\ 
\nonumber
&& k = \Bigl(M + \frac{\sum_{i=1}^3 L_i\,K_i-2\,L_4\,K_4}{2\,V}+
\frac{(K_1 K_2 -K_4^2)K_3}{V^2}\Bigr)\,(d\tau+A)+\omega\,,\\
&&*_3 d\omega = V\,dM-M\,dV +\frac{1}{2}\,\sum_{i=1}^3 (dL_i\,K_i
-L_i\,dK_i )- (dL_4\,K_4-L_4 \,dK_4)\,.
\nonumber
\eea
The harmonic functions that generate black ring solutions carrying 4
charges and 4 dipole charges are centred at a distance $R$ from the
origin of the 3D space~\eqref{3df}. The coordinates from this
special point will be indicated with $\Sigma$ and $\theta_\Sigma$
\be
\Sigma=\sqrt{r^2+R^2-2\,R\,r\,\cos\theta}\,,\quad \cos\theta_\Sigma = \frac{r\,\cos\theta-R}{\Sigma}\,.
\ee
The most general allowed harmonic functions that describe a ring
wrapping the fiber direction $\tau$ in~\eqref{ghm} are
\bea \label{brhf}
K_I = \frac{d_I}{\Sigma}\,,\quad L_I = \ell_I +\frac{Q_I}{\Sigma} \quad(I=1,\ldots ,4)\,,
\quad
M=m_0+\frac{m}{\Sigma}\,.
\eea

The parameters appearing in these harmonic functions should be chosen
so as to avoid Dirac-Misner singularities in the 1-form on
$R^3$, $\omega$. By focusing in a neighbourhood of the point
$r=0$ this requires
\be\label{DM0}
m_0 = - \frac{m}{R}\,,
\ee
while regularity around $\Sigma=0$ implies
\be\label{DMSigma}
m = \frac{R}{2}\,\Bigl(\sum_{i=i}^3 \ell_i\, d_i - 2 \,\ell_4\,d_4\Bigr)\,.
\ee
Note that condition (\ref{DM0}) also guarantees that the coefficient
of $(d\tau+A)$ in \eqref{f2t} (which we call $\mu$) vanishes at the
origin of polar coordinates $r=0$, and hence that $k$ is regular as a
1-form on $R^4$.

In order to insure the absence of closed time-like curves, the $Z_I$
and the coefficient of $d\tau^2$ should be positive. The conditions on
$Z_I$ are obviously satisfied if we take $\ell_i$, $Q_i$ and $d_i$, with
$i=1,2,3$, positive and $d_1 d_2 -d_4^2>0$.  Further conditions that
are sufficient to insure the positivity of $\alpha$ are
\bea
&&\ell_1\,\ell_2 - \ell_4^2>0\,,\quad \ell_2\,Q_1 + \ell_1\,Q_2-2\,\ell_4\,Q_4\ge 0\,,\quad \ell_1\,d_1+\ell_2\,d_2-2\,\ell_4\,d_4\ge 0\,,\nonumber\\
&&d_1\, Q_1 + d_2\, Q_2 - 2\,d_4\,Q_4\ge 0\,,\quad Q_1\,Q_2-Q_4^2\ge 0\,.
\eea
To analyze the condition on the $d\tau^2$ coefficient it is convenient
to rewrite the 5D 
metric in the first line of~\eqref{mm}
\bea
ds^2_E =  -\Bigl(\frac{\alpha}{Z_1 Z_2 Z_3}\Bigr)^{2/3}\,(dt+k)^2 + \Bigl(\frac{\alpha}{Z_1 Z_2 Z_3}\Bigr)^{-1/3}\,ds^2_4\,,
\eea
by completing squares with respect to $d\tau$. Then we obtain
\bea
ds^2_E &=& \Bigl(\frac{\alpha}{Z_1 Z_2 Z_3}\Bigr)^{2/3}\,\frac{\tilde I_4}{V^2}\Bigl[d\tau+A -\frac{\mu\,V^2}{\tilde I_4}\,(dt+\omega)\Bigr]^2\nonumber\\
&&+\Bigl(\frac{\alpha}{Z_1 Z_2 Z_3}\Bigr)^{-1/3}\,\frac{V}{\tilde I_4}\,[\tilde I_4\,ds^2_3 - (dt+\omega)^2]\,,
\eea
where
\bea\label{4ioc}
\tilde I_4 &=& \alpha^{-1}\,Z_1 Z_2 Z_3\,V^2-\mu^2\,V^2 \nonumber\\
&=&\frac{1}{2}\sum_{I<J=1}^3K_I\,K_J\,L_I\,L_J-\frac{1}{4}\sum_{I=1}^3 K_I^2\,L_I^2+V\,(L_1 L_2 - L_4^2)\,L_3\nonumber\\
&&+(K_1\,L_1+K_2\,L_2-K_3\,L_3)\,K_4\,L_4-K_4^2\,L_1\,L_2-K_1\,K_2\,L_4^2\nonumber\\
&&-2\,M\,(K_1\,K_2-K_4^2)\,K_3-M\,V\,(\sum_{I=1}^3 K_I\,L_I - 2\,K_4\,L_4)-M^2\,V^2\,.
\eea
Absence of closed time-like curves requires that $\tilde I_4$ be
everywhere positive. Then around $\Sigma=0$ we must have
\be
\tilde I_4 \approx  \frac{\tilde J_4}{\Sigma^4} > 0 \,,
\ee
where
\bea\label{tildeJ4}
\tilde J_4 &=&\frac{1}{2}\sum_{I<J=1}^3d_I\,d_J\,Q_I\,Q_J-\frac{1}{4}\sum_{I=1}^3 d_I^2\,Q_I^2\nonumber\\
&&+(d_1\,Q_1+d_2\,Q_2-d_3\,Q_3)\,d_4\,Q_4-d_4^2\,Q_1\,Q_2-d_1\,d_2\,Q_4^2\nonumber\\
&&-2\,m\,(d_1\,d_2-d_4^2)\,d_3\,,
\eea
and thus one needs $\tilde J_4>0$. 

We will now show that the geometry has a regular horizon of finite
area at $\Sigma=0$. The behavior of the metric functions for
$\Sigma\to 0$ is
\bea
&&Z_1 \approx \frac{R\,d_2\,d_3}{\Sigma^2}\,,\quad Z_2 \approx \frac{R\,d_1\,d_3}{\Sigma^2}\,,\quad Z_3 \approx \frac{R\,(d_1\,d_2-d_4^2)}{\Sigma^2}\,,\nonumber\\
&&Z_4\approx \frac{R\,d_3\,d_4}{\Sigma^2}\,,\quad \alpha^{-1}\approx \frac{d_1\,d_2-d_4^2}{d_1\,d_2}\,,\nonumber\\
&&\mu\approx \frac{R^2\,(d_1\,d_2-d_4^2)\,d_3}{\Sigma^3}\,,\quad \omega= O(\Sigma)\,,\quad \tilde I_4 \approx  \frac{\tilde J_4}{\Sigma^4}\,.
\eea
Looking at the metric in M-theory frame, one sees that the $T^6$ part
of the metric is finite, as far as $d_1 \,d_2\not = d_4^2$.  The gauge
fields $A_I$ ($I=1,\ldots,4$) have a potential divergence proportional
to $d\tau+A$; the divergent terms however cancel, thanks to the
identities
\bea
&&-\frac{\alpha\,\mu}{Z_1} + \frac{K_1}{V} = O(\Sigma^0)\,,\quad -\frac{\alpha\,\mu}{Z_2} + \frac{K_2}{V} =  O(\Sigma^0)\,,\nonumber\\
&&-\frac{\mu}{Z_3} + \frac{K_3}{V} = O(\Sigma^0)\,,\quad -\frac{\alpha\,Z_4\,\mu}{Z_1 Z_2} + \frac{K_4}{V}  = O(\Sigma^0)\,.
\eea
The term $\Bigl(\frac{\alpha}{Z_1 Z_2 Z_3}\Bigr)^{2/3}\,\frac{\tilde
  I_4}{V^2}$ that appears in $ds^2_E$ is finite and the term
$\frac{\mu\,V^2}{\tilde I_4}$ goes to zero. Finally the determinant of
the horizon metric (i.e. of the submanifold $t=\mathrm{const}$,
$\Sigma=0$) is
\be
\mathrm{det} g_\mathrm{Hor} = \Sigma^2 \,(\tilde I_4\,\Sigma^2\,\sin^2\theta_\Sigma-\omega^2_\phi)\approx \tilde J_4\,\sin^2\theta_\Sigma\,,\
\ee
so that the area of the horizon is proportional to $\tilde J_4^{1/2}$.

With the standard 5D-4D map~\cite{4D5D}, we
can use the isometry along $\tau$ and reduce the black ring
solution~\eqref{brhf} to a four dimensional black hole. So we expect
that the combination $\tilde I_4$ can be expressed in terms of the
quartic invariant ${\cal I}$ of the symmetry group E$_7$. This
invariant can be expressed in terms of the matrices
\be
Z_{AB} = \sum_{\hat{g},{h}=1}^8 (x_{\hat{g}{h}} + i
y_{\hat{g}\hat{h}}) (\gamma^{\hat{g}\hat{h}})_{AB}\,,~~~~  
Z^{AB} = \sum_{\hat{g},\hat{h}=1}^8 (x_{\hat{g}\hat{h}} + i
y_{\hat{g}\hat{h}}) (\gamma^{\hat{g}\hat{h}})^{AB}\,, 
\ee
where the $\gamma^g$'s are the $8 \times 8$ blocks of the $SO(8)$
Gamma matrices with chiral indices $A,B=1,\ldots,8$. Then, by
following ~\cite{Cremmer:1979up}, the quartic invariant can be written
as
\be\label{4ig}
{\cal J} = \frac{1}{256} \left[Z_{AB} Z^{BC} Z_{CD} Z^{DA} -
  \frac{1}{4} \left(Z_{AB} Z^{BA}\right)^2 + {\rm Pf}(Z_{AB})  + {\rm
    Pf}(Z^{AB})\right]\, 
\ee
where Pf indicates the Pfaffian of $Z$. As suggested by the expression
for the cycles related to our ansatz~\eqref{cycles}, one can use the
dictionary~\eqref{x2qd} and check that the general expression for the
quartic invariant~\eqref{4ig} agrees with the result in~\eqref{4ioc}
\begin{align}\label{x2qd}
&
x_{12} \to Q_1\,,~~
x_{34} \to Q_2\,,~~ 
x_{56} \to Q_3\,,~~
x_{13} \to Q_4\,,~~
x_{24} \to Q_4 \,,
\\ \nonumber
& 
y_{12} \to d_1\,,~~
y_{34} \to d_2\,,~~ 
y_{56} \to d_3\,,~~ 
y_{13} \to -d_4\,,~~
y_{24} \to -d_4\,,~~
y_{78} \to -2m \,. &
\end{align}

The 5D mass is proportional to
\be
M = \ell_2\,\ell_3\,\widetilde{Q}_{1}+\ell_1\,\ell_3\,\widetilde{Q}_{2}+(\ell_1\,\ell_2-\ell_4^2)\,\widetilde{Q}_{3}-2\ell_3\,\ell_4\,\widetilde{Q}_{4}\,,
\ee
where $\widetilde{Q}_I$ ($I=1,\ldots,4$) are the four charges carried by
the solution:
\be
\widetilde{Q}_1= Q_1+d_2\,d_3\,,\quad  \widetilde{Q}_{2}=Q_2+d_1\,d_3\,,\quad \widetilde{Q}_3 =  Q_3+d_1\,d_2-d_4^2\,,\quad \widetilde{Q}_4 = Q_4+d_3\,d_4\,.
\ee
In the IIB duality frame $\widetilde{Q}_1$, $\widetilde{Q}_2$,
$\widetilde{Q}_3$ correspond, respectively, to D1, D5 and momentum
charges; moreover there is an equal amount of F1 and NS5 charge, given
by $\widetilde{Q}_4$.

The angular momenta with respect to the 5D Cartan angles $\tilde\psi$
and $\tilde\phi$ (defined as $\tau = \tilde\psi+\tilde\phi$, $\phi
=\tilde\psi-\tilde\phi$) are given by
\bea
J_{\tilde\psi} &=& \frac{1}{2}\,\Bigl(\sum_{i=1}^3 d_i\,\tilde Q_i -2\,d_4\,\tilde Q_4\Bigr)-\frac{1}{2}\,(d_1\,d_2-d_4^2)\,d_3 + R\,\Bigl(\sum_{i=1}^3 \ell_i\,d_i - 2\,\ell_4\,d_4\Bigr)\,,\nonumber\\
J_{\tilde\phi} &=& \frac{1}{2}\,\Bigl(\sum_{i=1}^3 d_i\,\tilde Q_i -2\,d_4\,\tilde Q_4\Bigr)-\frac{1}{2}\,(d_1\,d_2-d_4^2)\,d_3 \,.
\eea

\section{Discussion and open problems} \label{daop}

We found an exact supersymmetric solution of type IIB supergravity
comprising all the fields that are sourced by the class of D1, D5 and
P bound states studied in~\cite{Giusto:2011fy}. The ansatz we find is
more general than the one that was previously used to describe
three-charge bound state geometries~\cite{standardBPS} 
and that, when
reduced to 5D, is equivalent to the STU model (i.e. $\mathcal{N}=2$ 5D
supergravity with two vector multiplets). We show that a suitable
chain of dualities relates our ansatz to the M-theory lift of
$\mathcal{N}=2$ 5D supergravity with three vector multiplets. 

We have also examined a black ring solution in this extended ansatz,
carrying four charges and four dipole charges. Though this solution is already
contained in the general class of solutions of 5D supergravity constructed 
in~\cite{BR}, an
explicit 11D embedding of this black ring solution 
is, as far as we
know, new. Moreover, we relate this M-theory solution
to a duality-equivalent IIB solution. Knowing the
solution in type IIB frame allows one to investigate the existence of
`near-horizon' limits containing an $AdS_3$ factor, and hence to
attempt a microscopic interpretation of the solution. As explained for
example in \cite{Bena:2004tk}, the near-horizon solution is obtained
by taking
\bea
r\sim \alpha'^2\,,\quad R\sim\alpha'^2\,,\quad Q_{1,2,4}\sim\alpha'\,,\quad Q_3\sim \alpha'^2\,,\quad d_{1,2,4}\sim\alpha'\,,\quad d_3\sim 1
\eea
and by sending $\alpha'\to 0$. This amounts to dropping the constants
$\ell_I$ from $L_1$, $L_2$ and $L_4$, but not from $L_3$. To obtain
the asymptotic limit of the near-horizon solution one then sends
$r\equiv \frac{\rho^2}{4} \to\infty$ and is left with the geometry:
\be
ds^2 = \frac{\bar\alpha\,\rho^2}{4\,\sqrt{\tilde Q_1\,\tilde Q_2}}(-dt^2+dy^2) + 4\,\sqrt{\tilde Q_1\tilde Q_2}\,\frac{d\rho^2}{\rho^2} + 4\,\sqrt{\tilde Q_1\tilde Q_2}\,d\Omega_3^2 + \sqrt{\frac{\tilde Q_1}{\tilde Q_2}}\,ds^2_{T^4}\,,
\ee
with
\be
\bar\alpha = \Bigl(1-\frac{\tilde Q_4^2}{\tilde Q_1\,\tilde Q_2}\Bigr)^{-1}\,.
\ee
This is $AdS_3\times S^3\times T^4$, where the radius of $AdS_3$ is
$\ell_{AdS} = 2\,(\tilde Q_1\,\tilde Q_2)^{1/4}$. The dilaton goes to
the constant value
\be
{\rm e}^{2\phi} = \bar\alpha\,\frac{\tilde Q_1}{\tilde Q_2}\,,
\ee
and thus, the usual computation gives for the central charge
\be\label{centralcharge}
c = \frac{3\,\ell_{AdS}}{2\,G_3} = \frac{6\,V_4}{\pi^4\,g^2\,\alpha'^4}\,\tilde Q_1\,\tilde Q_2\,\bar\alpha^{-1}\,,
\ee
where $V_4$ is the volume of $T^4$. To attempt a microscopic
interpretation one should express $c$ in terms of the integer charge
numbers; for the STU charges, this is done for example in
\cite{Bena:2004tk} (note that the charges $Q_I$ in~\cite{Bena:2004tk} are 4
times\footnote{This is due to the factor of $4$ between the 3D radial
coordinate $r$ and the 4D one $\rho$.} our $\tilde Q_I$):
\be
\tilde Q_1 = \frac{4\,\pi^4\,g\,\alpha'^3}{V_4}\,N_{D1}\,,\quad \tilde Q_2 = \frac{g\,\alpha'}{4}\,N_{D5}\,,\quad \tilde Q_3 = \frac{4\pi^4\,g^2\,\alpha'^4}{V_4\,R_y^2}\,N_P\,,
\ee
where $R_y$ is the radius of the $y$ circle and $N_{D1}$, $N_{D5}$, $N_P$ are the numbers of D1-branes, D5-branes and quantized units of momentum along $y$. The quantity $\tilde Q_4$ represents both the charge of fundamental strings wrapped on $y$ and of NS5-branes wrapped on $y$ and $T^4$, and hence it is quantized as
\be
\tilde Q_4 =  \frac{4\,\pi^4\,g^2\,\alpha'^3}{V_4}\,N_{F1}=  \frac{\alpha'}{4}\,N_{NS5}\,.
\ee  
Note that this imposes the constraint that the moduli combination $\frac{V_4}{(2\pi)^4\,g^2\,\alpha'^2}$ be a rational number. When we substitute these relations in (\ref{centralcharge}), we find
\be\label{centralchargebis}
c = 6\,(N_{D1}\,N_{D5}-N_{F1}\,N_{NS5})\,.
\ee

The CFT with central charge (\ref{centralchargebis}) that is dual to
our general black ring is not known, and it would be interesting to
further investigate its properties. One can however look at the subset
of black ring solutions with $\tilde Q_4=N_{F1}=N_{NS5}=0$ (but
$d_4\not=0$): it follows from our analysis that these are new black
ring solutions that should be describable within the same D1-D5 CFT,
with central charge $c= 6\,N_{D1} N_{D5}$, that describes the
Strominger-Vafa and the BMPV black holes. A consistent amount of work
\cite{Bena:2004tk,microscopicBR}
has been devoted to identifying the subset of states of the D1-D5 CFT
that correspond to the STU model black rings, but a completely
satisfactory understanding of this problem is still lacking. We point
out here that the same CFT should also contain a subset of states that
are dual to our more general class of black ring solutions:
characterizing these states and reproducing the entropy given in
(\ref{tildeJ4}) for $\tilde Q_4=0$ by counting them remains an
interesting open problem. Moreover in \cite{Bena:2011zw} a new
supersymmetric phase of the D1-D5 CFT that has no known gravity dual
was found: it would be interesting to see if our generalized black
rings could be relevant in that context.

\noindent {\large \textbf{Acknowledgements} }

\vspace{2mm} We thank I.~Bena, G.~Dall'Agata, S.~El-Showk, S.~Ferrara, M.~Gra\~na, S.~Mathur, J.F.~Morales, M.~Shigemori, H.~Triendl, D.~Turton, B.~Vercnocke and N.~Warner for several
enlightening discussions.  RR has been partially supported by STFC
Standard Grant ST/J000469/1 ``String Theory, Gauge Theory and Duality".

\appendix 

\section{Supergravity conventions} \label{sec:dualities}
\subsection{IIB}
The IIB action in string frame is
\bea
\Gamma &=& \int {\rm e}^{-2\phi}\,\Bigl(\sqrt{-g}\,R+4\, * d\phi \wedge d\phi -\frac{1}{2}\,*H^{(3)}\wedge H^{(3)}\Bigr)-\frac{1}{2}\,*F^{(1)}\wedge F^{(1)}\nonumber\\
&&-\frac{1}{2}\,*F^{(3)}\wedge F^{(3)}-\frac{1}{4} *F^{(5)}\wedge F^{(5)}+\frac{1}{2}\, H^{(3)}\wedge F^{(3)}\wedge C^{(4)}\,,
\eea
where
\be\label{fieldstrengths}
H^{(3)}= d B^{(2)}\,,\quad F^{(p+1)} = dC^{(p)}-H^{(3)}\wedge C^{(p-2)}\,,
\ee
and we define the Hodge dual in $d$ dimensions as
\be
* (dx^{i_1}\wedge \ldots \wedge dx^{i_p}) = \sqrt{-g}\,dx^{i_{p+1}}\wedge \ldots \wedge dx^{i_d}\,{\epsilon_{i_{p+1}\ldots i_{d}}}^{i_1\ldots i_p}\,,
\ee
with the orientation $\epsilon_{tyx_1x_2x_3x_4z_1z_2 z_3 z_4}=1$.
The 5-forms $F^{(5)}$ is taken to be self-dual.

We use the same supersymmetry conventions as in appendix A of \cite{Giusto:2011fy}.

To perform T-dualities we find convenient to work in the democratic formalism, and hence introduce the field strengths $F^{(7)}$ and $F^{(9)}$, dual to $F^{(3)}$ and $F^{(1)}$, as
\be\label{dualfields}
F^{(7)} = - * F^{(3)}\,,\quad F^{(9)} = * F^{(1)}\,.
\ee
The corresponding gauge potentials $C^{(6)}$ and $C^{(8)}$ are related to the field strengths by 
(\ref{fieldstrengths}).

The S-duality symmetry of IIB acts as
\bea
&&ds^2\to  {\rm e}^{-\phi}\,ds^2\,,\quad \phi\to  -\phi\,,\quad C^{(0)}\to C^{(0)}\,,\quad B^{(2)}\to  C^{(2)}\,,\quad C^{(2)}\to -B^{(2)}\,,\nonumber\\
&&C^{(4)}\to C^{(4)}-B^{(2)}\wedge C^{(2)}\,.
\eea
Note that this transformation leaves invariant the Einstein metric $e^{-\phi/2}\,ds^2$ and the 5-form $F^{(5)}$.
\subsection{T-duality}
Denote by $y$ the direction along which one performs T-duality and by $x^\mu$ the remaining coordinates. It is convenient to write the string metric, B-field and gauge fields as
\bea
ds^2 &=& G_{yy} \,(dy+A_\mu dx^\mu)^2 + \hat{g}_{\mu\nu} \,dx^\mu\, dx^\nu\,,\nonumber\\
B^{(2)}&=& B_{\mu y} \,dx^\mu \wedge (dy+A_\mu dx^\mu) + \hat{B}^{(2)}\,,\nonumber\\
C^{(p)}&=&C_y^{(p-1)}\wedge  (dy+A_\mu dx^\mu) + \hat{C}^{(p)}\,,
\label{trule0}
\eea
where the forms $\hat{B}^{(2)}$, $C_y^{(p-1)}$ and $\hat{F}^{(p)}$ are along the $x^\mu$ directions.

The T-duality transformed fields are
\bea
d{\tilde s}^2&=& G^{-1}_{yy} \,(dy-B_{\mu y} dx^\mu)^2 + \hat{g}_{\mu\nu}\, dx^\mu \,dx^\nu\,,\quad e^{2 \tilde \Phi}={e^{2\Phi}\over G_{yy}}\,,\nonumber\\
{\tilde B}^{(2)}&=&  - A_{\mu}\, dx^\mu dy + \hat{B}^{(2)}\,,\nonumber\\
{\tilde C}^{(p)}&=& \hat{C}^{(p-1)}\wedge(dy-B_{\mu y} dx^\mu)+C_y^{(p)}\,.
\label{trule}
\eea

\subsection{M-theory}
The M-theory action is
\be
\Gamma_{11} = \int \sqrt{-g_{11}}\,R_{11} -\frac{1}{2}\,*_{11} dA^{(3)} \wedge dA^{(3)} +\frac{1}{6}\,A^{(3)}\wedge dA^{(3)}\wedge d A^{(3)}\,,
\ee
where $R_{11}$ and $*_{11}$ denote the Ricci scalar and Hodge dual with respect to the 11D metric $g_{11}$. We use the orientation $\epsilon_{tx_1x_2x_3x_4 z_1 z_2 z_3 z_4 y z}=1$.

\bibliographystyle{utphys}      
\providecommand{\href}[2]{#2}\begingroup\raggedright\endgroup

\end{document}